\documentclass{article}
\usepackage{spconf,amsmath,graphicx}

\usepackage{amssymb,graphicx,amsmath,epstopdf,booktabs,array,subfigure,setspace,adjustbox,multirow,multicol,algorithm,algpseudocode,setspace,amsthm,pgfplots}
\usepackage{mathtools,tikz,url}
\usepackage{enumitem}

\DeclareMathOperator*{\argmin}{argmin}

\title{Conv-codes: Audio Hashing For Bird Species Classification}
\name{Anshul Thakur$^1$, Pulkit Sharma$^2$, Vinayak Abrol$^3$, Padmanabhan Rajan$^1$}
\address{
  $^1$School of Computing and Electrical Engineering, IIT Mandi, India\\
  $^2$Computational Health Informatics Lab, University of Oxford, UK \\ 
  $^3$Idiap Research Institute, Martigny, Switzerland\\
  Email: anshul\_thakur@students.iitmandi.ac.in}

\begin{document}
\ninept
\maketitle
\vspace{-0.2cm}
\begin{abstract}
In this work, we propose a supervised, convex representation based audio hashing framework for bird species classification. The proposed framework utilizes archetypal analysis, a matrix factorization technique, to obtain convex-sparse representations of a bird vocalization. 
These convex representations are hashed using Bloom filters with non-cryptographic hash functions to obtain compact binary codes, designated as conv-codes. The conv-codes extracted from the training examples are clustered using class-specific k-medoids clustering with Jaccard coefficient as the similarity metric. A hash table is populated using the cluster centers as keys while hash values/slots are pointers to the species identification information. During testing, the hash table is searched to find the species information corresponding to a cluster center that exhibits maximum similarity with the test conv-code. 
Hence, the proposed framework classifies a bird vocalization in the conv-code space and requires no explicit classifier or reconstruction error calculations. Apart from that, based on min-hash and direct addressing, we also propose a variant of the proposed framework that provides faster and effective classification. The performances of both these frameworks are compared with existing bird species classification frameworks on the audio recordings of 50 different bird species. 
\end{abstract}
\vspace{-0.25cm}
\section{Introduction}
\label{sec:intro}
\vspace{-0.15cm}
Acoustic monitoring is a suitable method to monitor and survey avian populations \cite{brandes2008automated,Lee} in their natural habitats. Bird species classification is an important module in such acoustic monitoring systems and directly helps in determining the avian diversity of an area of interest. 
One of the major problems in designing such a bird classification framework is the limited availability of labelled training data. 
Therefore, it may not be possible to utilize the data intensive classification frameworks such as deep neural networks (DNN) and convolutional neural networks (CNN) to their full potential. 
Hence, there is a need to develop classification techniques which could perform well even in scarcity of labelled training data.

Leveraging recent advances in dictionary learning based hashing for nearest neighbour retrieval \cite{compact_codes,robust_hash}, we propose archetypal analysis (AA) based supervised, convex-sparse hashing for species classification. The proposed hashing framework is an alternative to existing audio classification techniques and can easily be extended to other purposes such as audio indexing. The framework is characterized by the utilization of a hashing mechanism for classification and does not require any explicit classifier or reconstruction error calculations used in various dictionary learning frameworks. Generally, hashing based nearest neighbour search can be seen as a variant of template matching where hash codes act as templates for the train/test examples. However, the use of class-specific archetypal dictionaries provides generative modelling characteristics to the proposed framework. Hence, the proposed framework can be seen as a blend of generative modelling and hash-based template matching. Hashing based approaches such as \cite{shazam} have been successful in music information retrieval. However, these approaches cannot handle the large with-in class variabilities which are quite common in bioacoustic signals. The generative nature of the proposed framework helps in effectively overcoming this issue.     

The proposed hashing framework builds upon the work in \cite{compact_codes,ccse} and utilizes archetypal analysis (AA) \cite{CVPR_AA} for acoustic modelling. AA requires less amount of training data in comparison to other modelling techniques, such as Gaussian mixture models, for providing effective acoustic modelling \cite{aa_BAD}. The compressed super-frame (CSF) representation \cite{ccse} of bird vocalizations, obtained from the spectrogram (by embedding context to each frame), is given as input to the proposed framework. Each of the input CSF is hashed into a compact binary code designated as \emph{conv-codes} using class-specific AA and Bloom filters \cite{bloom} (see Section \ref{sec:prop}). These conv-codes are clustered in a class-specific manner using K-medoids with Jaccard coefficient \cite{jac} as similarity metric. The cluster centers obtained from this procedure are used for populating the hash table. The proposed framework utilizes the similarity between test conv-codes and these cluster centers for classification. Moreover, both conv-codes and cluster centers are bit strings, hence efficient bit-level operations can be used to calculate the aforementioned similarity. In this work, Jaccard coefficient \cite{jac} is used as a similarity measure for classification.    

In addition, this study also proposes a min-hash \cite{min} variant of the proposed framework. This variant approximates the Jaccard coefficient between two conv-codes using min-hashes extracted from convex-sparse representations. The utilization of min-hashing makes it possible to use the direct addressing \cite{direct} instead of searching the hash table for classification (see Section \ref{ssec:min} for details). The direct addressing reduces the hash table search to a constant time ($O(1)$), increasing the computational efficiency of the proposed framework.


The rest of the paper is organized as follows: 
Section \ref{sec:back} discuss some of the methods proposed in the literature for bird species classification using acoustic data. 
In Section \ref{sec:prop}, the proposed framework and its min-hash variant is described in detail. 
Performance analysis and conclusion are provided in Sections \ref{sec:perf} and \ref{sec:con}, respectively.

\vspace{-0.25cm} 
\section{Related Works}
\label{sec:back}
\vspace{-0.15cm}
Various methods targeting bird species classification have been proposed in the literature. 
In some classical methods, the syllables (the basic unit of birdsong) modelled as frequency and amplitude modulated sinusoidal pulses are used for identifying bird species \cite{Harma,Somervuo,syllable_pair}. 
However, the use of sinusoidal modelling based classification is limited to the species producing tonal sounds. Other methods based on support vector machines (SVM), deep neural networks (DNN) and convolutional neural networks (CNN) have also been proposed for this task. 
In \cite{Raich}, a convolutional neural network (CNN), having Segnet \cite{segnet} like architecture, is proposed to segment vocalizations from the spectrogram and simultaneously classifies these bird vocalizations. 
Hence, this framework overcomes the problem of segmentation of bird vocalizations before classification. 
However, this network requires pixel-wise labelling of spectrograms, which is generally not available. An unsupervised and scalable feature learning method based on spherical K-means is proposed in \cite{dan_skmeans}. 
These unsupervised features are shown to provide good classification performance for a large number of bird species, using a random forest classifier. 
In addition, SVM with various dynamic kernels such as probabilistic sequence kernel (PSK), GMM supervector (GMMSV) kernel, GMM-UBM mean interval (GUMI) kernel, GMM-based intermediate matching kernel (GMM-IMK) and GMM-based pyramid match kernel (PMK) have also been used for bird species classification \cite{deep}. 
In a recent study \cite{ccse}, AA is utilized to obtain the convex representations for bird species classification. This framework shows comparable performance to the existing classification frameworks including DNN and SVM.  

\begin{figure*}[t]
	\centering
	\includegraphics[scale=0.43]{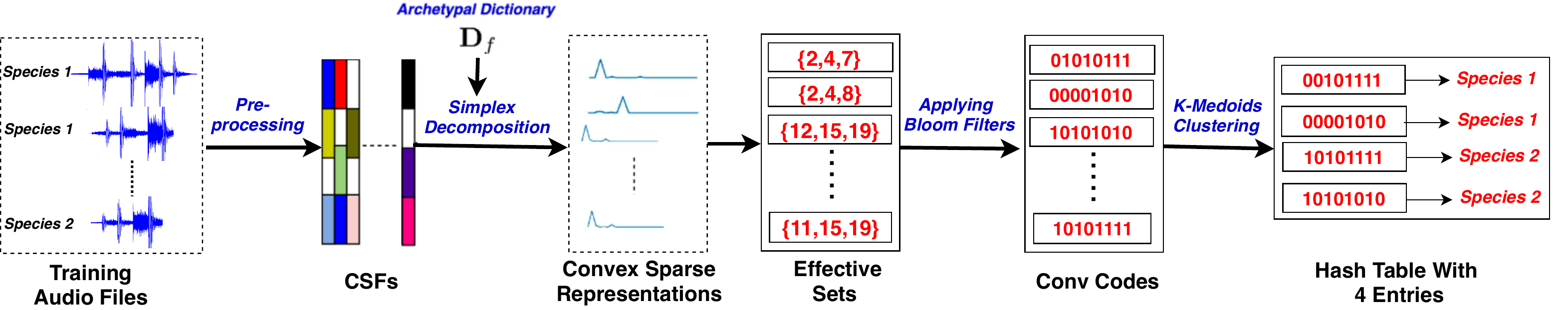}
	\caption{Illustration of the proposed convex sparse hashing approach for two
species. A dictionary of 20 archetypes is used for obtaining
convex representations. First ten archetypes ((1-10) correspond to
species 1 while remaining ten archetypes (11-20) correspond to species
2.
The cluster centers obtained by class-specific K-medoids clustering
(T=2) are used to obtain the hash table having 4 entries (2 per
species).}
	\label{fig:prop}
\end{figure*}
\vspace{-0.35cm}
\section{Proposed Framework}
\label{sec:prop}
\vspace{-0.15cm}
In this section, we describe the convex-sparse audio hashing in detail. First, we explain the process for obtaining archetypal dictionaries and the hash table. Then, the process to classify a given audio recording is described.
\vspace{-0.25cm}
\subsection{Preprocessing and dictionary learning}
\vspace{-0.1cm}
The audio recordings are converted into compressed super-frame (CSF) \cite{ccse} representation by concatenating $W$ neighbouring frames of the spectrogram and projecting the concatenated vector on a random Gaussian matrix. The concatenation helps in effectively capturing the frequency-temporal modulations that give species-specific nature to bird vocalizations, while random projections help in compressing the concatenated vectors ($Wm$-dimensional to $K$-dimensional, where $m$ is the number of frequency bins in a frame and $K<<Wm$). More details about CSF representation can be found in \cite{ccse}. Only CSFs corresponding to the vocalization regions of an audio are used for learning dictionaries. 
A semi-supervised segmentation method proposed in \cite{thakur_mlsp} is used here for segmenting bird vocalizations.

In this work, archetypal analysis (AA) \cite{CVPR_AA} is employed to obtain the class-specific dictionaries from the CSF representation. The CSFs obtained from vocalizations of a species are pooled together to form a matrix $\mathbf{X} \in \mathbb{R}^{K \times l}$, where $K$ is the dimensionality of the CSFs and $l$ is the number of pooled CSFs. 
This matrix $\mathbf{X}$ is factorized as: $\mathbf{X=DA}$. 
The dictionary $\mathbf{D} \in \mathbb{R}^{K\times d}$ has $d$ archetypes that model the convex hull of the data and characterize the extremal rather than the average behaviour as provided by other acoustic modelling techniques such as Gaussian mixture models. These archetypes are the convex combination of input data points such that $\mathbf{D=XB}$. The archetypal dictionary, $\mathbf{D}$ can be obtained by solving the following optimization problem:
\vspace{-0.25cm}
\begin{equation}
\begin{aligned}
 \argmin_{\substack{\mathbf{B},\mathbf{A}\\ \mathbf{b}_j \in \Delta_l,\mathbf{a}_i \in \Delta_d}} &\Vert \mathbf{X}\mathbf{-DA} \Vert_{F}^2= \argmin_{\substack{\mathbf{B},\mathbf{A}\\ \mathbf{b}_j \in \Delta_l,\mathbf{a}_i \in \Delta_d}}\Vert \mathbf{X}\mathbf{-XBA} \Vert_{F}^2,\\
 \Delta_l \triangleq [\mathbf{b}_j\succeq 0, &\Vert\mathbf{b}_j\Vert_1=1], \Delta_d \triangleq [\mathbf{a}_i\succeq 0, \Vert\mathbf{a}_i\Vert_1=1].
\label{eq:aa}
\end{aligned}
\end{equation} 

Here $\mathbf{a}_i$ and $\mathbf{b}_j$ are columns of $\mathbf{A} \in \mathbb{R}^{d\times l}$ and $\mathbf{B} \in \mathbb{R}^{l\times d}$, respectively. This optimization objective can be solved using the block-coordinate descent scheme. More details about the implementation of AA can be found in \cite{CVPR_AA}. The final dictionary $\mathbf{D}_f$ is obtained by concatenating all the class-specific dictionaries: $\mathbf{D}_f=[\mathbf{D}_{1} \mathbf{D}_{2} \ldots\mathbf{D}_{q}]$, where $\mathbf{D}_q$ is the archetypal dictionary of the $q$th class.

\vspace{-0.3cm}
\subsection{Generating conv-codes and populating hash table}
\label{ssec:conv_codes}
\vspace{-0.1cm}

A convex sparse representation ($\mathbf{y}_i$) is obtained by projecting a CSF ($\mathbf{x}_i$) on a simplex corresponding to the dictionary, $\mathbf{D}_f \in \mathbb{R}^{K \times qd}$ ($qd$ is the number of archetypes from all classes):
\vspace{-0.35cm}

\begin{equation}
\mathbf{y}_i=\argmin_{\substack{\mathbf{y}_i\\ \mathbf{y}_i \in \Delta_{qd}}} \Vert \mathbf{x}_i\mathbf{-D}_f\mathbf{y}_i \Vert^2_{2} 
\label{eq:active_Set}
\end{equation}
where $\Delta_{qd} \triangleq [\mathbf{y}_i\succeq 0, \Vert\mathbf{y}_i\Vert_1=1]$.

In this work, the active-set algorithm proposed in \cite{CVPR_AA} is used to solve equation \ref{eq:active_Set}. The convexity constraints on the decomposition leads to sparsity such that only few coefficients of $\mathbf{y}_i$ are significant \cite{CVPR_AA}. The archetypes corresponding to these significant coefficients have maximum contribution in representing the CSF. The set formed by indices of the top $Z$ coefficients having maximum magnitudes in convex sparse representation  is termed as \emph{effective-set}. An effective-set is computed for each CSF extracted from training audio recordings. Ideally, a characteristic or fixed set of archetypes contribute to the possible effective-sets for vocalizations of each bird species. Hence, effective-sets obtained for different bird species should be different, making them a suitable representation for the species classification task.

The set operations, required to work with effective-sets, are usually computationally expensive. On the other hand, the bit-level operations or manipulations in binary strings are relatively computationally efficient.  
Hence, these effective sets are converted into compact binary strings or conv-codes using Bloom filters \cite{bloom}. Bloom filters are designed to facilitate the set membership queries in an efficient manner by storing the information about the presence/absence of given elements in compact binary strings. To obtain a conv-code for a given effective set using Bloom filters, initially all bits in conv-code are set to 0. Then, the non-cryptographic hash functions of Bloom filters hashes each element of an effective set to one of the locations of the conv-code, where the corresponding bit is flipped to 1, marking the presence of that element in the given effective-set. These hash functions are a sequence of bit-level operations applied on an input element to produce a digest that corresponds to one of the location on the conv-code.

In this work, conv-codes extracted from the training audio recordings are used for populating the hash table. The conv-codes of each class are clustered using K-medoids (with Jaccard coefficient as the similarity metric) to obtain $T$ cluster centers. In K-medoids clustering, an input data point is the medoid or the cluster center. Hence, the cluster centers obtained using K-medoids are subset of the training conv-codes. These cluster centers are used as keys, pointing to the respective species or class label, in the hash table. Thus, the hash table has $qT$ entries, where $q$ is the number of classes. It must be noted that $qT$ is significantly less than the number of input CSFs. Figure \ref{fig:prop} depicts the process of creating the hash table from training input audio recordings of two bird species. 

\vspace{-0.3cm}
\subsection{Classification}
\label{ssec:test}
\vspace{-0.1cm}
In this section, we discuss the classification strategy employed for a test audio recording. 
Initially, a test input audio recording is processed to segment the bird vocalizations regions from the background.
A segmented bird vocalization is represented by a set of CSFs, $\mathcal{X}=[\mathbf{x}^t_1\mathbf{x}^t_2\ldots \mathbf{x}^t_n \ldots \mathbf{x}^t_s]$, where $\mathbf{x}^t_n$ is the $n$th CSF in the test vocalization. 
A convex-sparse representation ($\mathbf{y}^t_n$) is obtained for CSF ($\mathbf{x}^t_n$) using equation \ref{eq:active_Set}. 
This CSF, $\mathbf{x}^t_n$, is represented by an effective-set obtained by choosing the indicies of the top $Z$ coefficients having maximum magnitudes in $\mathbf{y}^t_n$. 
Finally, $\mathbf{x}^t_n$ is converted into a conv-code ($\mathbf{C}^t_n$) using the bloom filter. 
During classification, this conv-code is matched with all the conv-codes used for populating the hash tables during training. The hash table entry that exhibits maximum similarity with $\mathbf{C}^t_n$ is regarded as its nearest neighbour and $\mathbf{C}^t_n$ is assigned the species label pointed by the respective hash table entry.
In this work, we have used Jaccard coefficient \cite{jac} as a metric for this matching. The Jaccard coefficient between the test conv-code and the hash table entry compute the extent of intersection between their corresponding effective-sets.
As discussed earlier, the possible effective-sets for each class consist of a particular set of archetypes and ideally, this set of archetypes is unique for each bird species. 
Hence, two effective-sets belonging to same species exhibits maximum intersection and their corresponding conv-codes exhibit maximum Jaccard similarity. 
  
Each test CSF in $\mathcal{X}$ is classified into one of the available bird classes. A voting rule is applied on these CSF wise decisions to get the final species label of the bird vocalization represented by $\mathcal{X}$.  

\vspace{-0.4cm}
\subsection{Reducing computation using min-hashing}
\label{ssec:min}
\vspace{-0.1cm}
The retrieval time required to find a best matched conv-code from the hash table is $O(qT)$, where $qT$ is the number of hash table entries. Although bit-manipulation operations are fast, the required computation to find the best matched conv-code can be reduced further. To decrease the retrieval time, we propose a variant of the proposed framework that utilizes min-hashing \cite{shrivastava2012fast,min} on the convex-sparse representations. To obtain the min-hash, convex-sparse is randomly permuted and the indices of $Z$ most significant coefficients in this permuted representations are noted. The first of these indices act as min-hash for the convex-sparse representation under processing. It has been shown in the literature that the probability of two sets having same min-hash is equal to the expected Jaccard similarity between these sets \cite{mining, shrivastava2012fast}. Hence, it can be inferred that two sets having high similarity are most likely to produce the same min-hash.

Based on this, we propose to use direct addressing \cite{direct} for creating a hash table whose keys are min-hashes and values are corresponding label information. An array (a data-structure) can be seen as over-simplified example of the direct addressing. The index of an array is a key and the value corresponding to this key can be obtained by directly accessing the memory corresponding to this index. In this work, min-hashes are equivalent to indices of array and value at each index is the corresponding species label. 
Since min-hashes are indices of coefficients of the convex-sparse representation, the size of this array is equal to the number of archetypes in the dictionary. For testing a CSF, min-hash is obtained for its convex representation. The hash table is directly accessed at the location of this min-hash to obtain the species label. This hash table look-up requires $O(1)$ computation only.      

\vspace{-0.4cm}
\section{Performance Analysis}
\label{sec:perf}
\vspace{-0.2cm}
\subsection{Dataset Used}
\vspace{-0.1cm}
The proposed framework is evaluated on a dataset\footnotemark \footnotetext{We are in process of making the dataset public.} containing audio recordings of 50 different bird species, obtained from three different sources. 
The recordings of 26 bird species were obtained from the Great Himalayan national park (GHNP), in north India. These recordings were used in \cite{deep} for performance comparison. The recordings of 7 bird species were obtained from the bird audio database maintained by the Art \& Science centre, UCLA \cite{data2}. 
The remaining 17 bird species audio recordings were obtained from the Macaulay Library \cite{mac}. 
All the recording used here are 16-bit mono, sampled at 44.1 kHz and are of durations varying from $15$ seconds to $3$ minutes. 
The information about these 50 species along with the total number of recordings and vocalizations per species is available at \url{https://goo.gl/z6UEQa}. 

\begin{figure*}[t]
	\centering
	\includegraphics[trim={0cm 6.5cm 1cm 0cm},clip,scale=0.42]{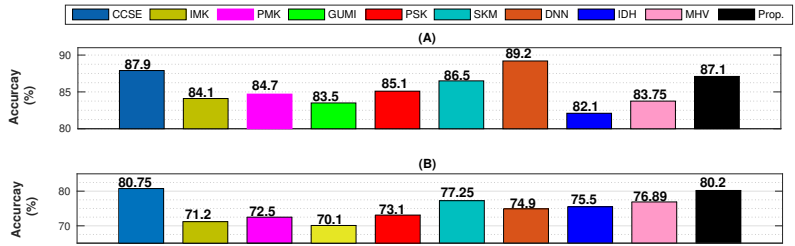}
	\caption{Classification performances of different methods obtained with (A) 33.33\% and (B) 20\% of the training data (Averaged across three and five folds respectively).}
	\label{fig:exp1}
\end{figure*}
\vspace{-0.3cm}
\subsection{Experimental setup}
\vspace{-0.1cm}
The spectrograms are obtained from each input audio recording by using the STFT with 512 FFT points on a frame rate of 20 ms with 50\% overlap. 
The window size $W=5$ is used to obtain super-frame representations which results in 1285-d ($1285=(257 \times 5)$) representation. 
Random projections are used to compress this 1285-d representation to a 500-d representation. 
A 25-atom archetypal dictionary is learned for each class and thus the final dictionary having 1250 (\textit{50}$\times $\textit{25}) atoms is used to obtain the convex sparse representations. 
The parameter $Z=4$ is used for obtaining effective-sets from these convex representations. 
A Bloom filter with 2 different non cryptographic hash functions i.e. Murmur3 \cite{murmur} and SpookyHash \cite{spooky} is used to convert the effective sets into 1024-bits conv-codes. K-medoids with 10 cluster centers (i.e. $T=10$) is applied on training conv-codes of each class to obtain the hash table entries. After training, the hash table containing 500 distinct entries ($10\times50$) is obtained.  All the parameters mentioned in this section are obtained empirically on a validation set.

\vspace{0.15cm}
\noindent\textbf{Comparative methods:} 
The classification performance of the proposed framework and its min-hash variant is compared with various existing bird species classification methods such as CCSE framework \cite{ccse}, SVM with dynamic kernels (IMK, PMK, GUMI and PSK), spherical K-means with random forest based framework proposed in \cite{dan_skmeans} and a DNN based approach proposed in \cite{deep}. 
CCSE framework is also based on super-frames and class-wise archetypal analysis, but it uses reconstruction error for classification. 
For SVM and DNN based classification schemes, MFCC using delta and acceleration coefficients are used as the feature representation. 
In addition, we have also compared the proposed framework with incoherent dictionary hashing (IDH) \cite{compact_codes} for the application of species classification, using nearest neighbour retrieval. 
For a fair comparison, an incoherent dictionary of 1250 atoms (same as the one used in the proposed framework) is used in IDH. Apart from that, the performance of the proposed framework is also compared with its min-hash variant (MHV).

\vspace{0.1cm}
\noindent\textbf{Train/test data distribution:}
The 20\% of the vocalizations present in the audio recordings of each class are used for the validation purposes. These vocalizations are not used for training or testing.
The remaining vocalizations are used for performance comparison using a three-fold cross validation scheme. In each fold, 33.33\% of the vocalizations (per class) are used for training while the remaining are used for testing. 
The results presented here are averaged across three folds. 
\vspace{-0.3cm}

\subsection{Results and Discussion}
\vspace{-0.1cm}
The classification performance of the proposed framework along with other comparative methods is shown in Fig.~\ref{fig:exp1}(A). 
Following inferences can be made from this figure: 
\begin{itemize}[leftmargin=*]
\item The classification performance of the proposed framework is comparable to the DNN i.e. DNN shows a relative improvement of only 2.1\%. 
\item Similar classification performances are observed for CCSE and the proposed framework, with CCSE showing a relative improvement of 0.9\% over the proposed approach. 
This observation indicates that maximum intersection between effective-sets of two super-frames is same as the minimum class/dictionary specific reconstruction error used for classification, as in the CCSE approach. 

\item The proposed framework outperforms IMK, PMK, GUMI, PSK and SKM by a relative improvement of 3.44\%, 2.76\%, 4.13\%, 2.3\% and 0.6\%, respectively.

\item The incoherent dictionary learning used in IDH could not perform up-to the level of AA used in the proposed framework. 
This is due to the fact that the proposed framework utilizes class-specific AA dictionaries which provide more discriminative representations.

\item Min-hash variant shows comparable performance to SVM based frameworks. However, on comparison with the proposed framework, min-hash variant shows a drop of 3.84\% in classification accuracy. As discussed earlier, this drop in classification accuracy occurs due to the loss of information incurred while representing effective-sets by one index i.e. min-hash only. However, using min-hashing decreases the required classification time by approximately three times.  

\end{itemize}
\vspace{-0.1cm}

Although the proposed method performs similar to CCSE framework, however it is computationally efficient. The minimum reconstruction error classification used in CCSE is computationally expensive in comparison to the indexing. In our experiments, it has been observed that average time taken for classifying a test bird vocalization using the proposed hashing framework is $0.13$ seconds as opposed to $0.4$ seconds in CCSE. Similarly, for the min-hash variant, this average classification time is $0.05$ seconds. Hence, min-hash variant makes up for the drop in accuracy by providing rapid classification. The running times mentioned here are the average of ten runs for classifying approximately $3200$ vocalizations.

\vspace{0.1cm}
\noindent\textbf{Performance in low data conditions:} To analyse the performance of the proposed framework and other comparative methods in low data conditions, we conducted the same experiment, with the same parameter setting but with different train/test data distribution. 
Here, we have used only 20\% of the available recordings per class for training and the rest for testing. 
In this experiment, a $5$-fold cross-validation is used for performance comparison. 
The results of this experiment are illustrated in Fig.~\ref{fig:exp1}(B). 
It can be observed that all the methods exhibits a drop in average performance when compared with the previous experiment (using 33.3\% recordings for the training). 
However, this drop is more significant for methods that uses classifiers such as DNN, SVM and random forest. 
The relative drops of 15.34\%, 14.4\%, 16.05\%, 14.1\%, 10.69\% and 16.03\% are observed for IMK, PMK, GUMI, PSK, SKM and DNN, respectively. 
On the contrary, dictionary learning methods (CCSE, IDH, the proposed framework and min-hash variant) observe a relatively low drop in average classification performance. 
The relative drops of 8.13\%, 8.04\%, 7.93\% and 8.18\% are observed for CCSE, IDH, the proposed framework and min-hash variant. This justifies the claim that the proposed framework requires less amount of training data to provide effective classification as compared to the SVM and DNN based methods.


\vspace{-0.5cm}
\section{Conclusion}
\label{sec:con}
\vspace{-0.17cm}
In this work, we proposed an AA based convex sparse hashing framework for bird species classification. 
Bloom filters are used to convert convex representations into binary conv-codes, that are further used for indexing and classification. 
The experiments demonstrate that the classification performance of the proposed framework is similar to state-of-the art methods. 
Our experimentation has validated that this framework requires less training data in comparison to the SVM and DNN based classification frameworks. 
No extra effort has been made to make the conv-codes more compact and to make the retrieval time more efficient. 
In future, we will work on this aspect to further decrease the time required for classification.

\bibliographystyle{IEEEbib}

\end{document}